\documentclass[final,3p,times,twocolumn]{elsarticle}

\usepackage[switch]{lineno}
\usepackage{amsmath,amssymb,amsfonts}
\usepackage{graphicx}
\usepackage{subfigure}
\usepackage{textcomp,nicefrac}

\usepackage{amssymb}

\usepackage{lineno}

\begin{document}
\renewcommand\floatpagefraction{.75}
\renewcommand\topfraction{.75}
\renewcommand\bottomfraction{.75}
\renewcommand\textfraction{.1}
\setcounter{totalnumber}{50}
\setcounter{topnumber}{50}
\setcounter{bottomnumber}{50}

\begin{frontmatter}



\title{
The Performance of AC-coupled Strip LGAD developed by IHEP}



\author[label1,label2]{Weiyi Sun}
\author[label1,label3]{Mengzhao Li \corref{cor1}}
\author[label1]{Zhijun Liang}
\author[label1]{Mei Zhao \corref{cor1}}
\author[label4]{Xiaoxu Zhang}
\author[label1,label2]{Tianyuan Zhang}
\author[label1,label2]{Yuan Feng}
\author[label1,label2]{Shuqi Li}
\author[label1,label2]{Xinhui Huang}
\author[label1]{Yunyun Fan}
\author[label1]{Tianya Wu}
\author[label1]{Xuan Yang}
\author[label1]{Bo Liu}
\author[label1]{Wei Wang}
\author[label1]{Yuekun Heng}
\author[label5]{Gaobo Xu}
\author[label1]{João Guimaraes da Costa}
\affiliation[label1]{organization={Institute of High Energy Physics, Chinese Academy of Sciences},
            addressline={19B Yuquan Road}, 
            city={Beijing},
            postcode={100049}, 
            country={China}}

\affiliation[label2]{organization={University of Chinese Academy of Sciences},
            addressline={19A Yuquan Road}, 
            city={Beijing},
            postcode={100049}, 
            country={China}}
\affiliation[label3]{organization={China Center of Advanced Science and Technology},
            addressline={55, Zhongguancun East Road}, 
            city={Beijing},
            postcode={100190}, 
            country={China}}

\affiliation[label4]{organization={School of Physics, Nanjing University},
            addressline={22 Hankou Road, Gulou District}, 
            city={Nanjing},
            postcode={210093}, 
            country={China}}
            
\affiliation[label5]{organization={Institute of Microelectronics, Chinese Academy of Sciences},
            addressline={3 Beitucheng West Road}, 
            city={Beijing},
            postcode={100029}, 
            country={China}}
\cortext[cor1]{Corresponding author: Mengzhao Li (mzli@ihep.ac.cn), Mei Zhao (zhaomei@ihep.ac.cn)}

\begin{abstract}
The AC-coupled Strip LGAD (Strip AC-LGAD) is a novel LGAD design that diminishes the density of readout electronics through the use of strip electrodes, enabling the simultaneous measurement of time and spatial information. The Institute of High Energy Physics has designed a long Strip AC-LGAD prototype with a strip electrode length of 5.7 mm and pitches of 150 $\mu m$, 200 $\mu m$, and 250 $\mu m$. Spatial and timing resolutions of the long Strip AC-LGAD are studied by pico-second laser test and beta source tests. 
 The laser test demonstrates that spatial resolution improves as the pitch size 
 decreases, with an optimal resolution achieved at 8.3 $\mu$m. Furthermore, the Beta source test yields a timing resolution of 37.6 ps. 
\end{abstract}


\end{frontmatter}



\section{Introduction}
\label{sec:introduction}
Low Gain Avalanche Diode (LGAD) have demonstrated excellent timing performance, making them essential for precise timing measurements in particle physics experiments\cite{Pellegrini_2014,Moffat2018}. However, LGAD's ability to achieve high-precision spatial resolution is limited due to the limitation of pixel size caused by dead zones (Junction Termination Extension (JTE) and P-stop)\cite{Giacomini2021}. To overcome this limitation, AC-Coupled LGADs (AC-LGADs) have been developed as an extension of LGAD technology\cite{Giacomini2019,Nakamura2021}. 
AC-LGADs offer the capability to simultaneously measure time and space information with high precision and provide a 100\% fill factor \cite{Kim2016,Giacomini2019}. The Strip AC-LGAD, a novel structure featuring AC-coupled LGAD with strip-shaped readout electrodes, offers the potential to decrease the readout electronic density compared to AC-LGAD with pixel electrodes\cite{Paternoster2017,ONARU2021164664}. Strip AC-LGAD has gathered significant interest and has been considered to be used in Electron-Ion Collider (EIC) simultaneously as trackers and time-of-flight detector and has the potential to be used in future colliders\cite{osti_1764596,Apresyan_2020,Wada_2019}. 

The Institute of High Energy Physics(IHEP) has designed a long Strip AC-LGAD prototype, which utilizes strip electrodes of a 5.7 mm length. In this study, we investigate the timing resolution and spatial resolution of long-strip AC-LGADs. To perform this analysis, laser and Beta tests were employed\cite{TORNAGO2021165319,9489306}. This study will help to optimize the design of AC-LGAD detectors in the future. 

\section{Introduction to Strip AC-LGAD}
\label{sec:design}
\subsection{Design of Strip AC-LGAD}
\begin{figure}[htbp]
\centering
\subfigure[]{
\includegraphics[width=0.8\columnwidth]{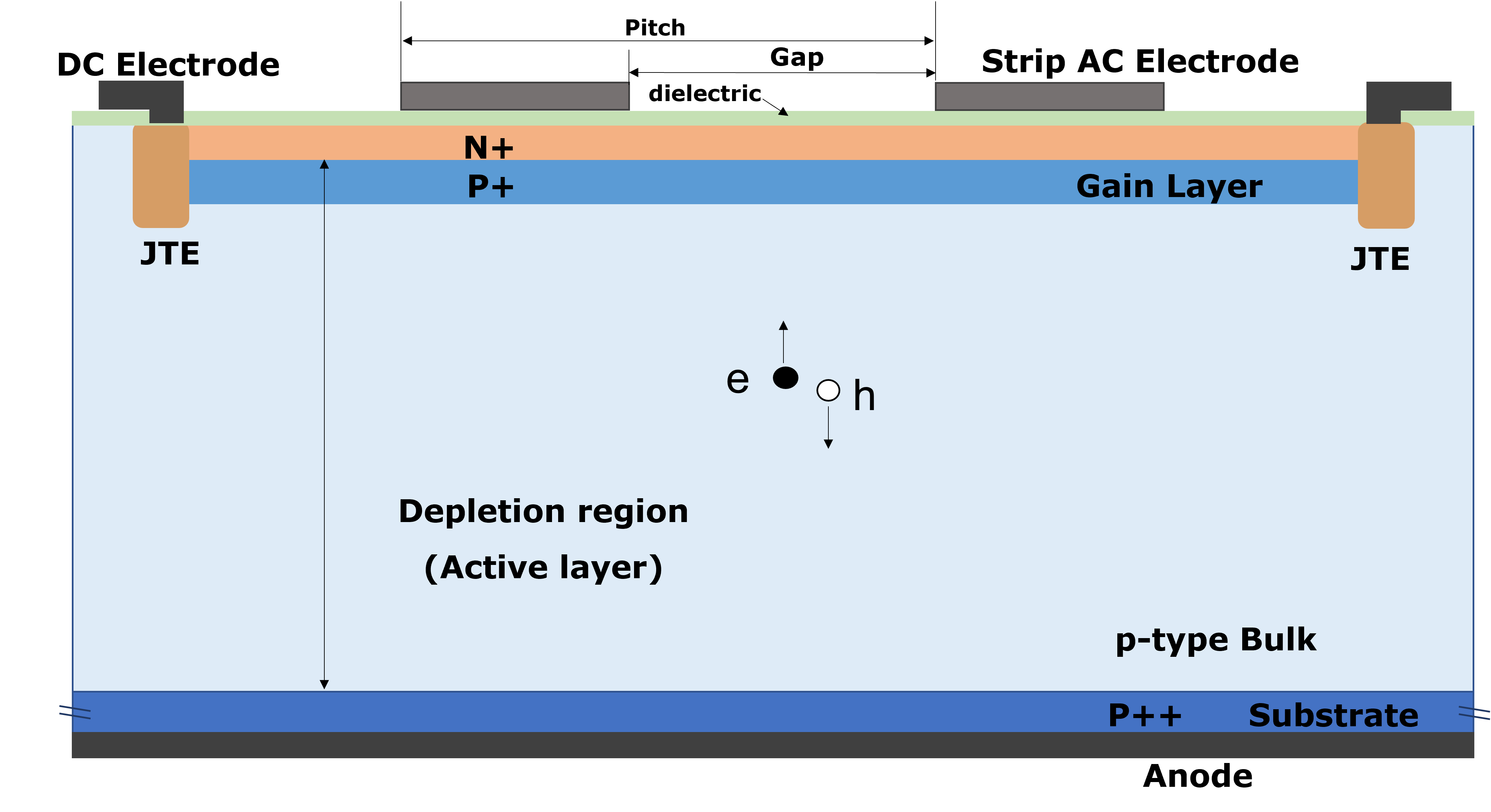}
}
\subfigure[]{ 
\includegraphics[width=0.8\columnwidth]{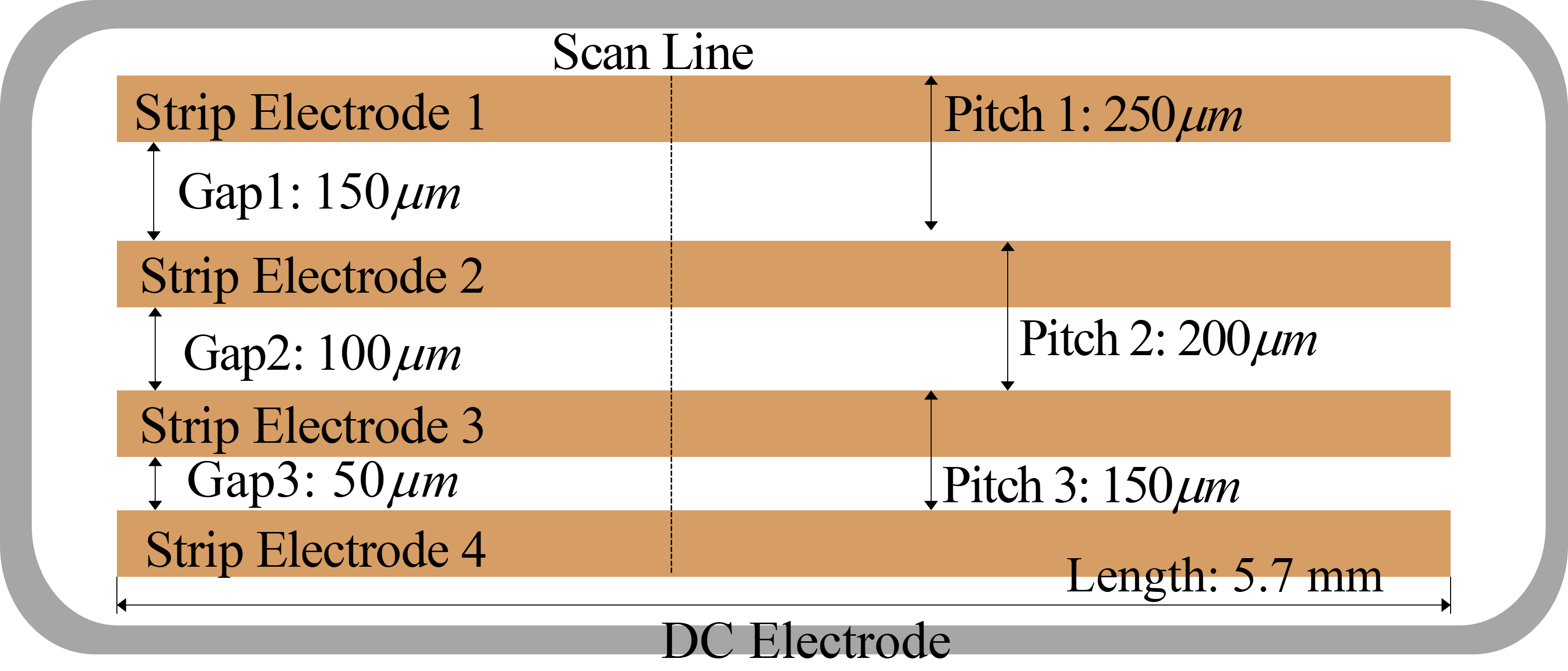}
}
\caption{Schematic of sectional view (a) and top view (b) of long Strip AC-LGAD. (Not to scale) The length of the strip electrodes is 5.7 mm.  The width of four strip electrodes is 100 $\mu m$. }
\label{structure}
\end{figure}

 The long Strip AC-LGAD sensor used in this experiment is designed by IHEP and fabricated by the Institute of Microelectronics (IME). The device contains strip electrodes, dielectric, DC electrode,  N$^{+}$ layer, P$^{+}$ layer, P-type bulk, P$^{++}$ layer, and aluminum anode as illustrated in \figurename~\ref{structure}(a).  
 The Strip AC-LGAD sensor has a 50 $\mu m$ p-type epitaxial layer (active layer). 
 The length of the strip electrodes is 5.7 mm.
 The schematic of the long Strip AC-LGAD is shown in \figurename~\ref{structure}(b). Three pitch sizes of  250 $\mu m$, 200 $\mu m$, and 150 $\mu m$ are designed with the widths of the strip electrodes being 100$\mu m$. The gaps between two adjacent strip electrodes are 150 $\mu m$, 100 $\mu m$, and 50 $\mu m$. 
\subsection{Capacitance properties of Strip AC-LGAD}

\begin{figure}[htbp]
\centering
\includegraphics[width=0.9\linewidth]{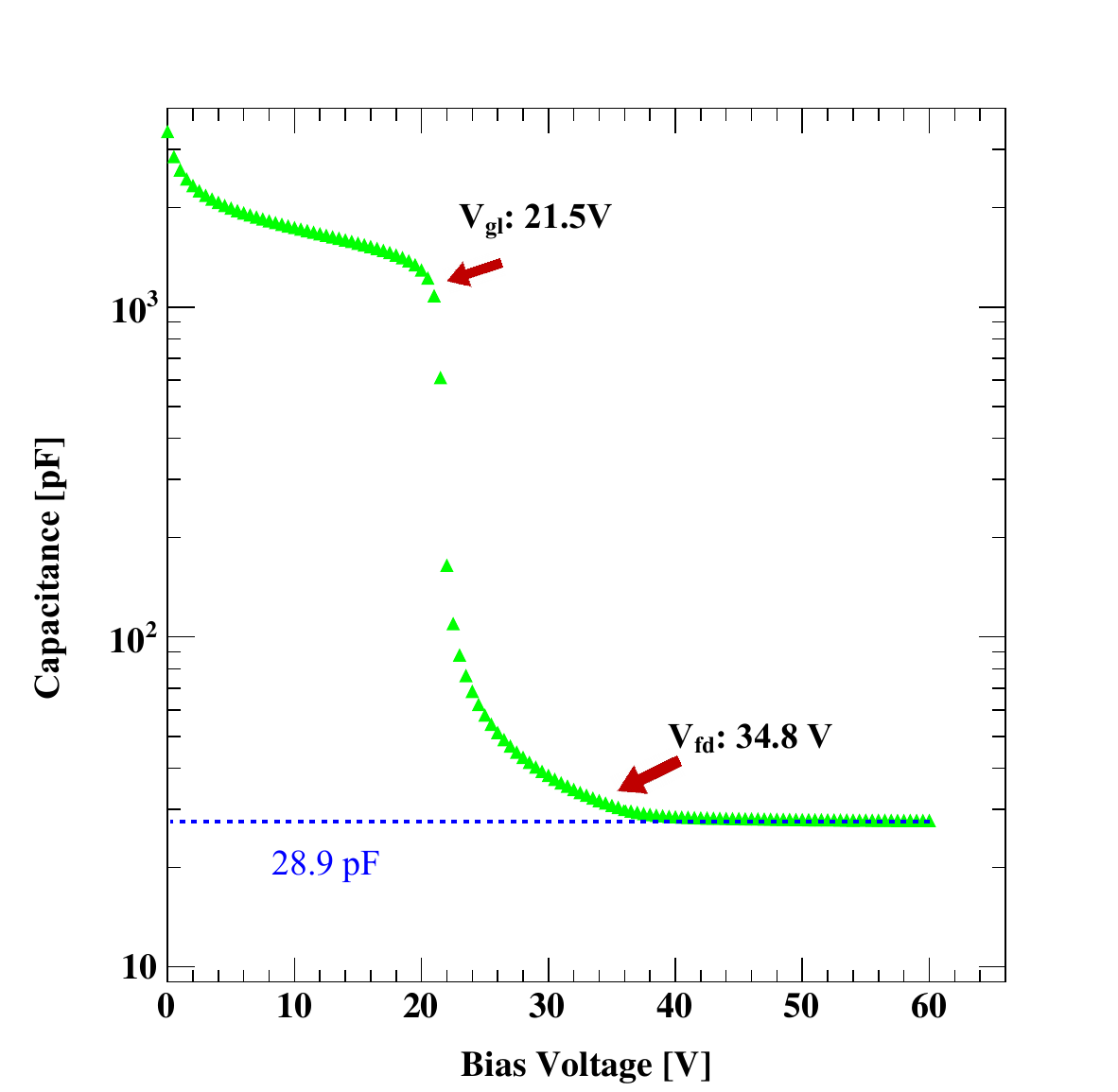}
\caption{C-V characteristic of Strip AC-LGAD. }
\label{cvl}
\end{figure}

The capacitance between the N+ layer and anode $vs.$ bias voltage is shown in \figurename~\ref{cvl}(a). The capacitance value is greater than 2000 pF when no bias is applied and 28.9 pF when fully depleted. 
 The gain layer depletion voltage $V_{gl}$ of strip AC-LGAD is 21.5 V, and the full depletion voltage $V_{fd}$ is 34.8 V. 
The capacitance between strip AC electrodes and the N+ layer is measured to be $170\pm0.2$ pF and does not change with the bias voltage up to 60V. 

\section{Experimental Setups}

\begin{figure}[t]
\centerline{\includegraphics[width=1.75in]{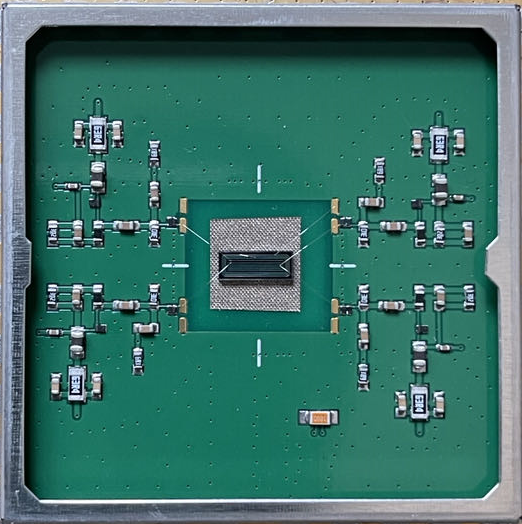}}
\caption{Strip electrodes of the long Strip AC-LGAD are bonded to the four-channel read-out board. }
\label{RDB}
\end{figure}

  Laser and Beta tests are used to investigate timing and spatial resolutions. The laser and Beta platforms contain read-out boards, amplifiers,  oscilloscopes, a laser source, or a Beta source, respectively. The long Strip AC-LGAD electrodes are bonded to the four-channel readout board, as shown in \figurename~\ref{RDB}. The four-channel readout board is designed and manufactured according to the University of California Santa Cruz (UCSC) single-channel readout board\cite{CARTIGLIA201783}.  The electrodes' signals are recorded by the oscilloscope (Teledyne LeCroy HDO9204) after the pre-amplifier and second amplifier. The oscilloscope's bandwidth is 2 GHz, and the sampling rate is 20 Gs/s for four channels.

\begin{figure}[htbp]
\centering
\subfigure[]{
\includegraphics[width=0.8\linewidth]{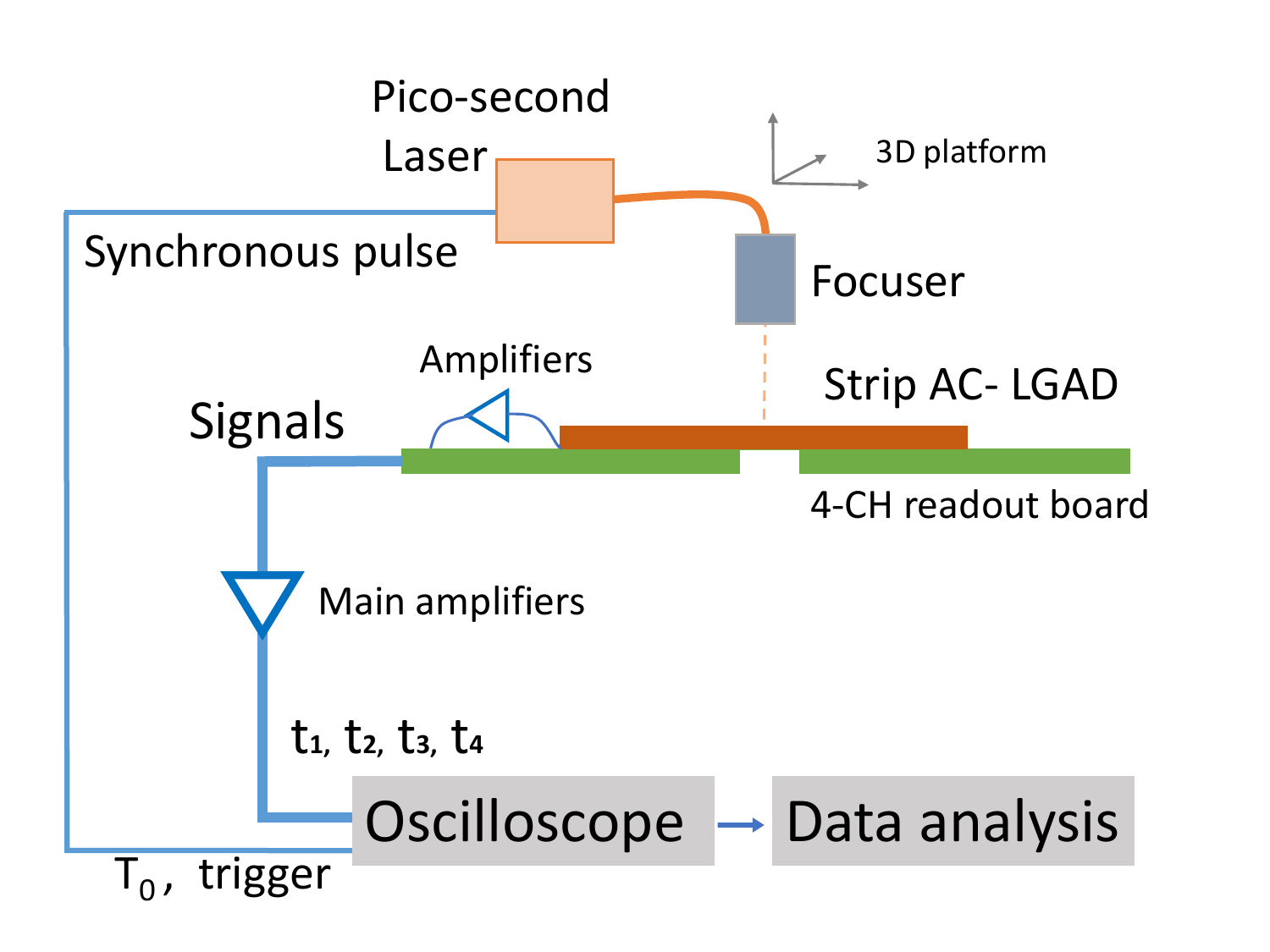}
}
\subfigure[]{
\includegraphics[width=0.8\linewidth]{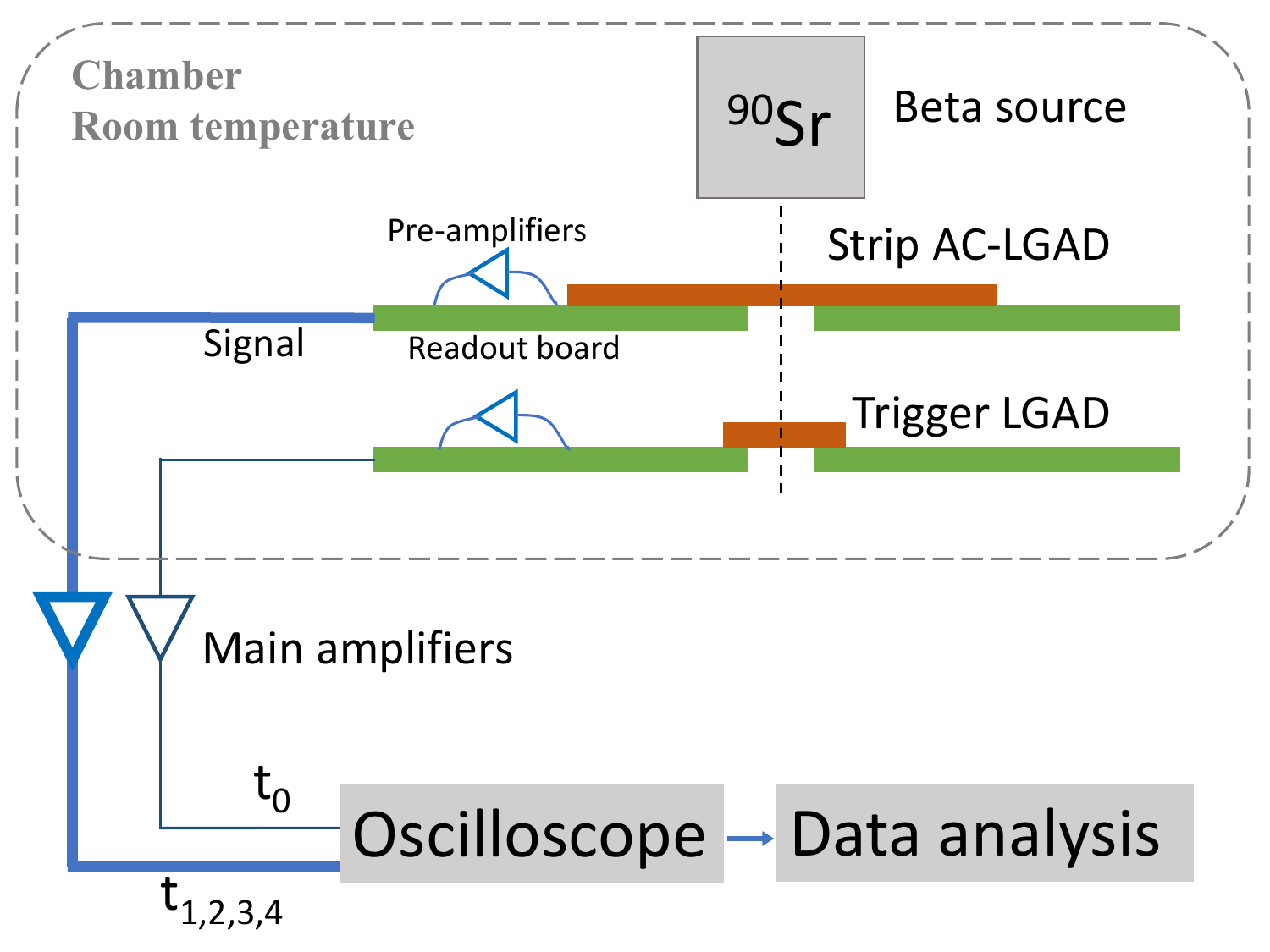}
}
\caption{The schematic of the laser platform (a) and Beta Platform (b).}
\label{TCT}
\end{figure}

The read-out board is fixed to the Beta and laser platforms, as shown in \figurename~\ref{TCT}.
For the laser test, a picosecond laser pulse with a wavelength of 1064 nm hit the Strip AC-LGAD at a frequency of 20 MHz through a focuser. The focuser is fixed on the pinning table to achieve high-precision movement better than 1 $\mu m$.  
 The laser test starts from the upper edge of strip electrode 1 and ends at the lower edge of strip electrode 4 with a step of 2 $\mu m$, as shown in \figurename~\ref{structure}(b). 1000 waveforms are recorded at each point before moving to the next point. Furthermore, the laser spot size remains within $1.5\ \mu m$. Since the width of strip electrodes is the same, gaps will be used to characterize pitch in the following.

For the Beta test, a calibrated LGAD (trigger LGAD) is placed beneath the Strip AC-LGAD as a trigger, and a $^{90}$Sr beta source is placed above the Strip AC-LGAD.

\section{Results of laser and beta tests}
\label{analysis}
\subsection{Laser test}
Laser hit positions are reconstructed, and thereby, spatial resolution is studied through laser tests. The position reconstruction method relies on the variation in signal amplitudes between the two strip electrodes closest to the hit position. When laser or particles hit gap 1, the signal amplitudes of electrodes 1 and 2 change with the hit position. The hit positions are characterized by the ratio $R$, defined as:
\begin{equation}
R=\frac{Amp_2}{Amp_1+Amp_2}
\end{equation}
Here, $Amp_1$ and $Amp_2$ represent the signal amplitudes of strip electrodes 1 and 2, respectively.

\begin{figure}[t]
\centering
\subfigure[]{
\includegraphics[width=0.8\linewidth]{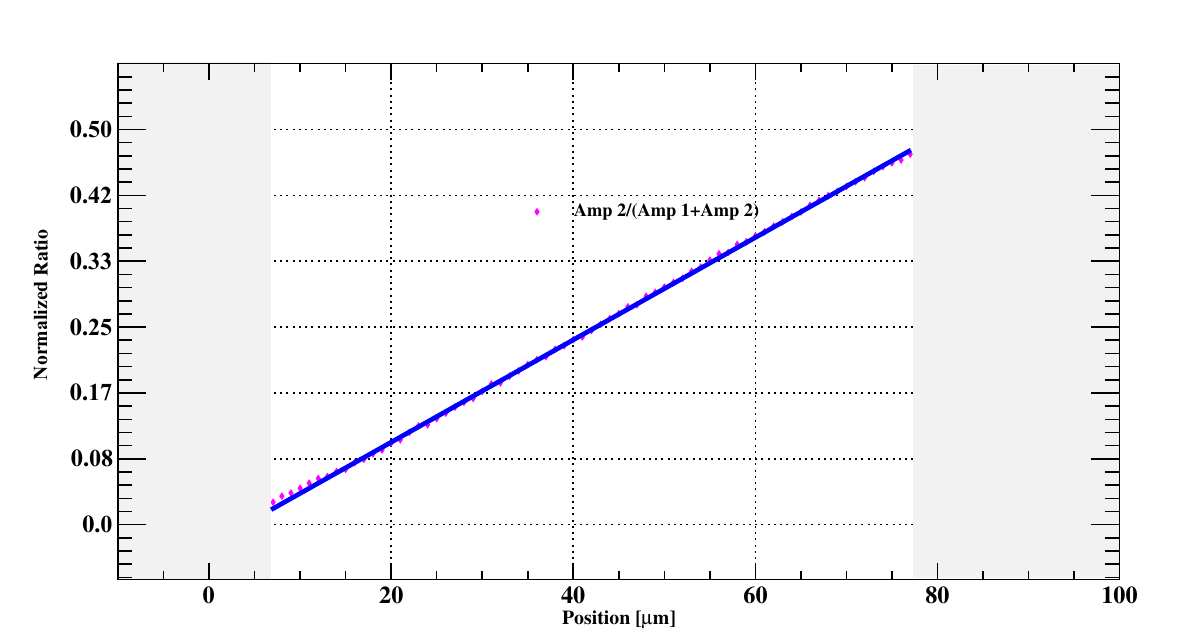}
}
\subfigure[]{
\includegraphics[width=0.8\linewidth]{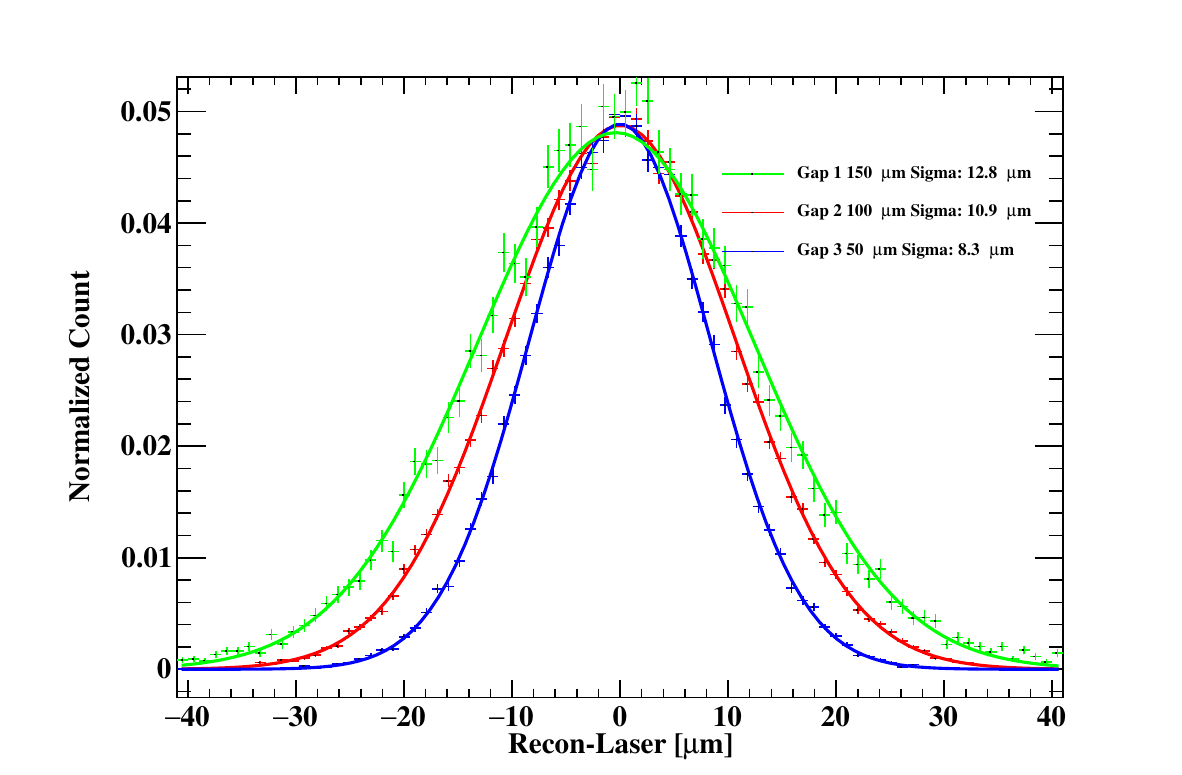}
}

\caption{Relationship between R and hit position, linear fit is applied.(a) Difference between laser hit position and the reconstructed position of three gaps with Gaussian fit. The resolutions of gap 1, gap 2, and gap 3 are 12.8 $\mu m$, 10.9 $\mu m$, and 8.3 $\mu m$, respectively.(b)}
\label{resl5}
\end{figure}

The relationship between $R$ and the hit position is approximately linear and can be fitted with $R=k_R\cdot x+c$, as shown in \figurename~\ref{resl5}(a). Therefore, the laser hit position $x$ can be calculated by the following equation:

\begin{equation}
x=\frac{R-c}{k_R}
\label{fit}
\end{equation}
Here, $R$ is the ratio, $k_R$, and $c$ are the fitted slope of $R$ and the fitted constant, respectively. 

The sigma of the difference between the laser and the reconstructed position is defined as spatial resolution.
To obtain the spatial resolution, 1000 waveforms per laser hit position are recorded, resulting in 1000 reconstructed positions per laser hit position. The difference between the reconstructed and laser hit positions of all hits in the gap is plotted as one histogram. The sigma of the fit corresponds to the spatial resolution of the gap. \figurename~\ref{resl5}(b) shows the histogram of the differences between the reconstructed and true positions. The spatial resolutions of gap 1, gap 2, and gap 3 are determined to be 12.8 $\mu m$, 10.9 $\mu m$, and 8.3 $\mu m$, respectively.

\subsection{Beta test}
\label{analysisbeta}
 The timing resolution of the whole strip AC-LGAD is defined as the sigma of the following distribution:
\begin{equation}
\Delta T=T_{trigger}-\frac{\Sigma_i a_i^2T_i}{\Sigma_i a_i^2}
\end{equation}
$T_{1},T_{2},T_{3}$, and $T_{trigger}$ are hit times measured on strip electrodes 1, 2, 3, and trigger LGAD.
Since the oscilloscope has only four channels, the signal of strip electrode 4 is not involved in the calculation of time resolution.
$a_1^2, a_2^2, a_3^2$ are signal amplitudes of strip electrodes 1, 2, and 3, and the weighting is to increase the contribution of the strip electrode with the largest signal amplitude to reduce the effect of noise on the timing resolution.
\begin{figure}[t]
\centerline{\includegraphics[width=0.8\columnwidth]{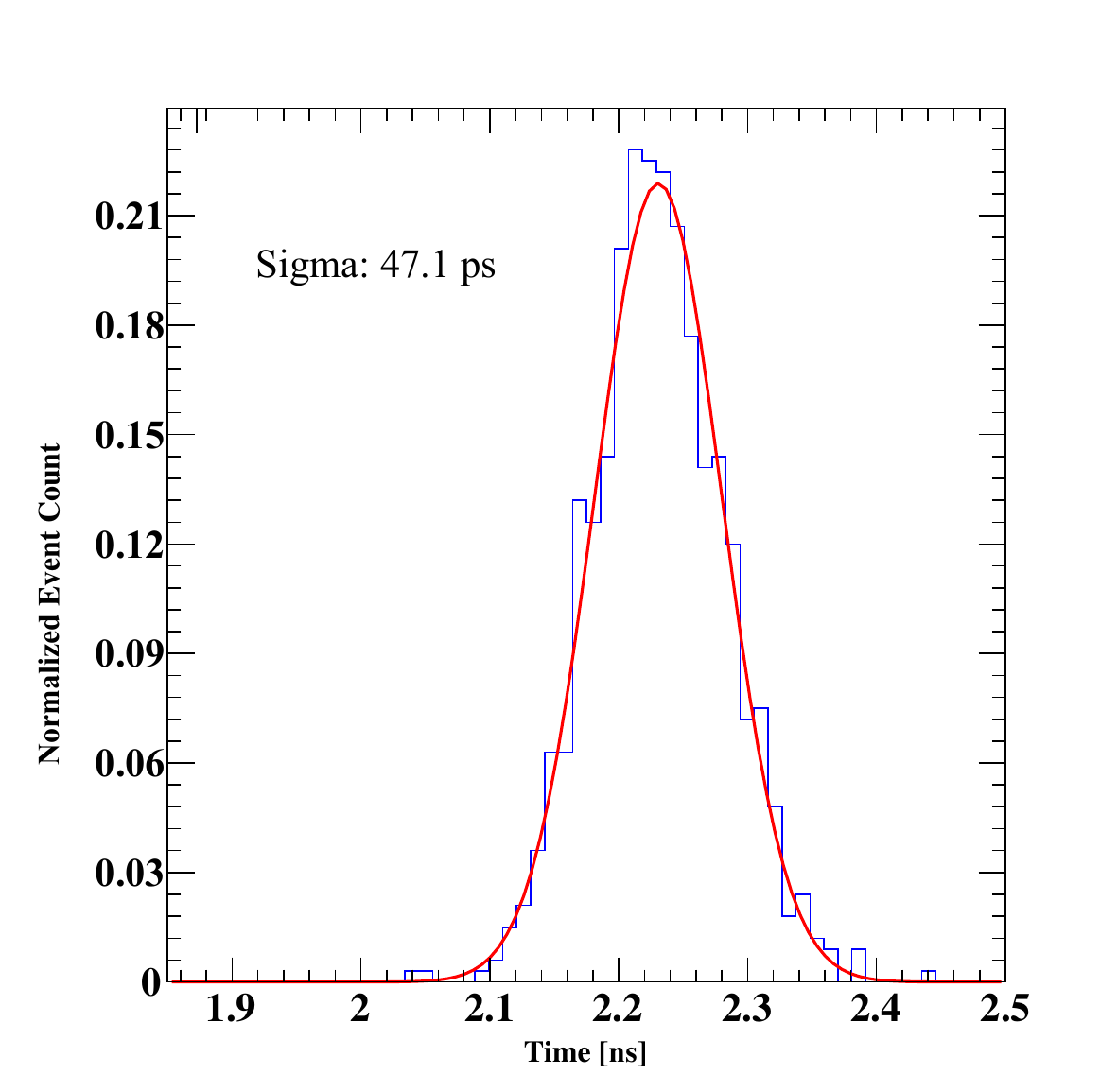}}
\caption{Distribution of $\Delta T$. }
\label{TE2}
\end{figure}
The distribution of $\Delta T$  is shown in  \figurename~\ref{TE2}. The timing resolution of trigger LGAD is 28.5 ps,  the sigma of the distribution is 47.1 ps, and the timing resolution of the long Strip AC-LGAD is 37.6 ps. \section{Conclusion}
The Strip AC-LGAD detectors, characterized by their lower readout electronic density, exhibit significant potential for application in future colliders. IHEP has designed a long Strip AC-LGAD prototype with a 5.7 mm length and three pitches and conducted laser and Beta tests to investigate the variations in spatial and timing resolutions of different gaps. 

Experimental results show that the timing resolution is about 37.6 ps, and the spatial resolutions of long Strip AC-LGAD with pitch sizes of 150 $\mu m$, 200 $\mu m$, and 250 $\mu m$ are 8.3 $\mu m$, 10.9 $\mu m$, 12.8 $\mu m$, respectively. The performance of the long Strip AC-LGAD shows that it is feasible for a 4D-tracker in future particle physics experiments.\section*{Acknowledgments}
This work was supported in part by the National Natural Science Foundation of China under Grant 12042507, Grant 12175252, Grant 12275290, Grant 11961141014, Grant 12105298; in part by the China Postdoctoral Science Foundation under Grant 2022M722964; in part by the National Key R$\&$D Program of China under Grant 2022YFE0116900; in part by the State Key Laboratory of Particle Detection and Electronics under Grant SKLPDE-ZZ-202315.

  \bibliographystyle{elsarticle-num} 
  \bibliography{AC_LGAD_REF.bib}



\end{document}